\documentclass[sigconf]{acmart}
\AtBeginDocument{%
  }

\renewcommand\footnotetextcopyrightpermission[1]{}

\setcopyright{none}

\usepackage{mathtools}
\usepackage{listings}
\usepackage{
  float,
  wrapfig,
  graphics,
  graphicx,
  subcaption
}

\usepackage{framed}

\usepackage{xspace} 
\usepackage{multirow} 
\usepackage{pifont}
\usepackage{hyperref}
\usepackage{colortbl} 

\newcommand{\CodeQL}{\textsc{CodeQL}\xspace}
\newcommand{\Gopher}{\textsc{Gopher}\xspace}
\newcommand{\Gosec}{\textsc{Gosec}\xspace}
\newcommand{\SnykCode}{\textsc{Snyk~Code}\xspace}

\newcommand\totanumberdetection{7,473\xspace} 
\newcommand\datasetsize{328\xspace}
\newcommand\totalnumberofrules{14\xspace}
\newcommand*{\lstref}[1]{\lstlistingname~\ref{#1}}
\newcommand*{\none}[0]{--}

\newcommand*{\na}[0]{{--}}

\definecolor{StrongRed}{HTML}{fac3c3}
\definecolor{TableGreen}{HTML}{bfe0af}
\definecolor{RoyalBlue}{HTML}{4169E1}
\definecolor{ForestGreen}{HTML}{228B22}
\definecolor{BrickRed}{HTML}{CB4154}

\colorlet{SoftRed}{StrongRed}  

\lstdefinelanguage{QL}{
  keywords={class, extends, exists, not, and, or, if, else, return, function},
  sensitive=true,
  comment=[l]{//},
  morecomment=[s]{/*}{*/},
  string=[b]",
  morestring=[b]'
}

\newenvironment{findingbox}{%
    \MakeFramed{%
        \hsize=0.95\columnwidth%
        \advance\hsize-2\fboxsep%
        \advance\hsize-2\fboxrule%
        \FrameRestore%
    }%
    \noindent\raggedright%
}{\endMakeFramed}

\begin{document}

\title{Evaluating Cryptographic API Misuse Detectors for Go}

\author{Vivi Andersson}
\email{vivia@kth.se}
\orcid{0009-0000-6519-625X}
\affiliation{%
  \institution{KTH Royal Institute of Technology}
  \city{Stockholm}
  \country{Sweden}
}

\author{Martin Monperrus}
\orcid{0000-0003-3505-3383}
\email{monperrus@kth.se}
\affiliation{%
  \institution{KTH Royal Institute of Technology}
  \city{Stockholm}
  \country{Sweden}
}
\renewcommand{\shortauthors}{Andersson and Monperrus}
\begin{abstract}
Cryptographic API misuse represents a critical vulnerability class that undermines the security foundations of modern software. Yet, it remains largely unexplored in Go despite its dominance in security-critical infrastructure. This paper presents the first comprehensive study of cryptographic API misuse detection in Go, identifying and analyzing 4 state-of-the-art tools (\CodeQL, \Gopher, \Gosec, and \SnykCode) and establishing a consolidated taxonomy of \totalnumberofrules relevant misuse classes. Through an experimental evaluation of 328 security-critical open-source Go projects, we discovered 7,473 cryptographic API misuses, providing insights into the prevalence and distribution of these vulnerabilities. Our systematic comparison reveals significant variations in misuse coverage, with immediate practical implications for security engineers and long-term implications for research in this domain.
\end{abstract}




\maketitle
\pagestyle{plain}
\thispagestyle{plain}
\section{Introduction}

The misuse of cryptographic APIs is a critical class of code vulnerabilities \cite{mosavi2023detecting}.
Cryptographic API misuses are caused by the essential complexity of cryptographic protocols and the specialized knowledge required to use them correctly \cite{afroseEvaluationStaticVulnerability2023, patnaikUsabilitySmellsAnalysis, frantzMethodsBenchmarkDetecting2024}. Yet, correct usage of these cryptographic APIs is required to maintain the security these protocols promise.
When cryptographic APIs are wrongly used, the major problem is that developers have a false sense of cryptographic security, founded on mathematical truth, while the actual software is weak and insecure.
For example, skipping certificate validation in TLS makes applications susceptible to man-in-the-middle attacks through nation-state firewalls \cite{ensafi2015analyzing}.

Cryptographic misuse detection has been extensively studied in the Java ecosystem \cite{chenPreciseReportingCryptographic2024, xiaExploringAutomaticCryptographic2024, wickertFixNotFix2023, piccolboni2021crylogger, rahamanCryptoGuardHighPrecision2019, krugerCogniCryptSupportingDevelopers2017,  nadiJumpingHoopsWhy2016}. Yet, Java is not the stack used for most security-critical modern software infrastructure. Today, Go, aka Golang, is a major stack for infrastructure software\footnote{\href{https://go.dev/wiki/GoUsers}{https://go.dev/wiki/GoUsers}}. 
It powers certificate authorities (e.g., Let's Encrypt's Boulder), service meshes (e.g., Traefik), container orchestration platforms (e.g., Kubernetes), and infrastructure-as-code tools (e.g., Terraform).
Despite Go's inevitability in security-critical systems, cryptographic misuse detection remains understudied. 
To the best of our knowledge, no prior work has systematically analyzed cryptographic misuse detection for Go.
We address this gap with the first qualitative and quantitative analysis.

Go has a comprehensive cryptographic API suite. For example, the Go library \verb|crypto/tls| provides support for the protocol TLS.
As in other ecosystems, Go developers remain prone to insecure code due to cryptographic API misuse  \cite{zhangGopherHighPrecisionDeepDive2024a, liCryptoGoAutomaticDetection2022}. Compared to Java, Go’s smaller standard library and more opinionated defaults can reduce certain algorithm‑selection errors, shifting failure modes toward configuration and usage mistakes.
Two recent, high-profile CVEs highlight these risks: CVE-2024-45337 in \texttt{golang.org/x/ssh}, where flawed assumptions about public-key authentication enabled authentication bypasses \cite{goBypassCVE}, and CVE-2025-66491 in Traefik, where \texttt{Verify=On} inverted verification semantics and disabled TLS certificate checks \cite{CVE-2025-66491}.

We define a \textit{cryptographic API misuse} as the incorrect usage of cryptographic functions through an API, which introduces security vulnerabilities and compromises the theoretical guarantees of the cryptographic functions used.
From there, we study the problem of cryptographic API misuse in Go as follows.
First, we conduct a survey of tools for detecting cryptographic API misuse in Go. Next, we analyze those tools (incl. the corresponding paper and documentation) in order to create a consolidated taxonomy of cryptographic API misuse for Go.
Second, we perform a quantitative experimental campaign. We collect security-relevant open-source projects in Go, run all the tools over them, and collect the detected API misuses. This enables us to study prevalence, tool consensus, and performance of misuse detection tools in Go.

\textbf{Results.} We study four state-of-the-art tools spanning academic, open-source, and commercial origins: \CodeQL, \Gopher, \Gosec, and \SnykCode. We consolidate \totalnumberofrules\ cryptographic misuse classes that (i) are relevant in Go and (ii) are supported by at least one tool; several have appeared in recent CVEs. Running all four tools on 328 open-source projects yields \totanumberdetection\ detected API misuses, enabling a systematic comparison across misuse classes. These findings help practitioners (e.g., Kubernetes security engineers) choose and interpret tools, acknowledging, for instance, that not all provide character-level precision, and highlight research needs, notably the formalization and enforcement of API-level security invariants. We also release a replication package to facilitate follow-up work.\footnote{\url{https://github.com/ASSERT-KTH/crypto-api-misuse-detectors-go}}

\textbf{Contributions.}
\begin{itemize}
\item The first systematic analysis of the crypto-misuse problem space in the Go infrastructure domain.
\item A large-scale experiment collecting \totanumberdetection\ misuses across \datasetsize\ notable open-source projects using four state-of-the-art detection tools.
\end{itemize}

\section{Experimental Methodology}
\subsection{Research Questions}
The objective of this research is to evaluate the state-of-the-art in detecting cryptographic API misuse in Go. We aim to identify and analyze the strengths and weaknesses of existing approaches, structuring the evaluation around two research questions:
\paragraph{RQ1. What is the design space of state-of-the-art tools for detecting cryptographic API misuses in Go?}
In this RQ, we aim to 1) identify the existing tools applicable to go, 2) which types of misuses they detect, and 3)  by which methods. We first identify relevant tools through a systematic literature search (\autoref{sec:tool-survey}), then analyze their technical design and detection capabilities. For academic tools, we extract information primarily from the corresponding paper.  
For industrial tools, we survey the official documentation.
We classify each tool based on its analysis approach and categorize the types of cryptographic vulnerabilities each tool detects.

\paragraph{RQ2. To what extent are crypto misuse detection tools for Go consistent with each other?}
In this RQ, we aim to quantitatively compare the concrete performance of the considered tools when applied to real-world Go projects. We start by assembling a representative dataset of Go modules that depend on cryptographic libraries for application security (\autoref{sec:dataset}). Each tool is then run over this dataset, and its detections are matched and compared.

\subsection{Considered Tools}\label{sec:tool-survey}
We conduct a systematic search to identify tools for detecting cryptographic API misuse in Go. We start by searching Google Scholar using the query \verb|'crypto*| \verb|API| \verb|misuse| \verb|detection| \verb|AND tool'|. We manually review each paper and confirm that it proposes a tool with Go support. 
Then, we conducted a broader web search targeting industry and community-developed tools. Since not all of them may use the terminology of ``misuse'', we focus the search query on the keyword \verb|crypto| \verb|vulnerability|.

We curate a systematic list of tools that meet the following inclusion criteria: (1) Support for Go, (2) Capability to detect cryptographic API misuse, (3) Available documentation, (4) Active development (e.g., a commit in the last 6 months).

Our search yields five tools, of which four meet our criteria. From academia, we identify \textit{\Gopher} \cite{zhangGopherHighPrecisionDeepDive2024a}, a specialized Go API misuse detection tool. One academic tool \textit{\textsc{CryptoGo}} \cite{liCryptoGoAutomaticDetection2022} is no longer available and is therefore excluded.
From the industry, we find three actively maintained tools with Go support. \textit{\CodeQL} \cite{CodeQLQueryHelp2025} is GitHub's Datalog-based framework with built-in Go analysis and rules for detecting cryptographic API issues. \textit{\Gosec} \cite{gosec} is a community-developed security checker for Go, and \textit{\SnykCode} \cite{GoSnykUser2025} is a proprietary, free-of-charge, static application testing tool from the company Snyk.
To the best of our knowledge, this is the most up-to-date and comprehensive list of tools to detect cryptographic API misuse in Go.

\subsection{Open-source Projects Dataset}\label{sec:dataset}
We base our quantitative evaluations on a dataset of real-world Go projects from Chen et al.~\cite{chenEmpiricalStudyCgo2025}. The original dataset includes 920 unique, highly-starred repositories collected via crawling GitHub and systematic filtering (removing duplicates, archived, inactive, and educational repositories). We further refine this dataset as follows.

\paragraph{Repository Health} We remove archived or unavailable modules and retain only well-maintained projects with commits in the past year and stable releases (indicated by non-pre-release Git tags).

\begin{figure}[h]
    \centering
    \includegraphics[width=\linewidth]{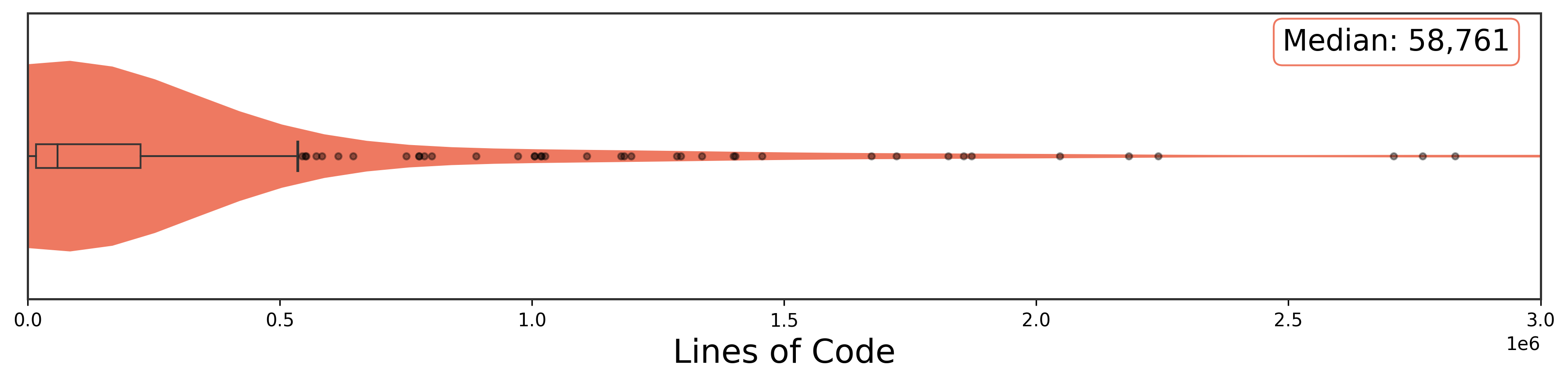}
    \caption{Lines of code per repository}
    \Description{A violin plot of Lines of Code for 328 repositories showing a strongly right-skewed distribution, where most repositories have relatively few lines of code and a small number of large outliers extend into the millions.}
    \label{fig:dataset_loc}
\end{figure}

\paragraph{Cryptographic Relevance}
We exclude projects that do not import any Go standard \verb|crypto| packages. Following prior work~\cite{chenPreciseReportingCryptographic2024, zhangAutomaticDetectionJava2023, wickertFixNotFix2023}, we manually assess each project's cryptographic security relevance, classifying project types and retaining only those depending on cryptography for core functionality. For example, \textit{Gops}\footnote{\href{https://github.com/google/gops}{https://github.com/google/gops}} is excluded since cryptography is not central to its purpose.
\paragraph{Descriptive Statistics} The final dataset comprises 328 security-relevant projects, ranging from 7 to over 8 million lines of code (median: 58,761)  (\autoref{fig:dataset_loc}). These actively maintained projects have a median of 9,096 stars and 167 contributors, reflecting strong community adoption.

\subsection{Protocol for RQ1}
We conduct a systematic qualitative comparison of the 4 tools identified in \autoref{sec:tool-survey}: \CodeQL, \Gopher, \Gosec, and \SnykCode. We compare the detection coverage across tools and their technical implementation.

\textbf{Taxonomy Rules. }To enable systematic comparison across tools with divergent coverage and classification approaches, we establish a unified taxonomy of cryptographic misuse issues. Each rule in our taxonomy represents a distinct type of cryptographic API misuse that can be detected through static analysis. We begin by collecting all distinct detection rules from each tool, preserving their original granularity and categorization. This forms a baseline of available issues detected across all evaluated tools. We then merge overlapping rules into common categories. 
For example, \CodeQL uses a single rule ``CWE-327'' for detecting TLS issues, while \Gopher represents these as 2 distinct rules. We merge these into a unified TLS security category. This consolidation creates a comparable taxonomy across all evaluated tools. We map each issue to security advisories where available.

We retain each rule's original severity classification (High, Medium, Low) as reported by the source. For each tool, we assess which of the rules is supported. We do so by surveying the accompanying paper and official documentation.

\subsection{Protocol for RQ2}\label{sec:rq2-method}
We evaluate the tools' practical detection capabilities by running each tool on our dataset and comparing their detections.

\newcolumntype{C}[1]{>{\centering\arraybackslash}p{#1}}
\begin{table*}[t]
  \caption{Detection of cryptographic misuses by tools for Golang, per their documentation.}
    \begin{minipage}[t][1.2\baselineskip]{0.55\textwidth}
      \small\hspace{30mm}\textbf{(a)} Taxonomy of Cryptographic API Misuses in Go.
    \end{minipage}%
    \hfill
    \begin{minipage}[t][1.2\baselineskip]{0.35\textwidth}
      \small\textbf{(b)} Detection support across tools.
    \end{minipage}
  \label{tab:taxonomy-rules}
  \centering
  \footnotesize
  \renewcommand{\arraystretch}{0.92}
  \begin{tabular}{p{1.9cm}C{2mm}p{2.6cm}C{6mm}l|C{1cm}C{1cm}C{1cm}C{1cm}}
    \toprule
    \textbf{Category} & \textbf{ID}& \textbf{Issue} & \textbf{Sev.} & \textbf{Advisory}& \textbf{CodeQL} & \textbf{Gopher} & \textbf{Gosec} & \textbf{Snyk} \\
    \midrule
  \multirow{3}{*}{\parbox{2.2cm}{Cryptographic\\Primitives}} 
    & 01 & Insecure algorithms & High & \href{https://www.cve.org/CVERecord?id=CVE-2024-55885}{CVE-2024-55885} &  & \checkmark & \checkmark & \checkmark \\
    & 02& Crypto insecure PRNG & Med & \href{https://nvd.nist.gov/vuln/detail/CVE-2024-21495}{CVE-2024-21495} & \checkmark & \checkmark & \checkmark & \checkmark \\
    & 03&  Deprecated Go function & Low & \na{} &  & \checkmark & & \\
    \midrule
    \multirow{3}{*}{\parbox{2.2cm}{Key Management}} 
     &04 & Constant/predictable key & High & \href{https://www.cve.org/CVERecord?id=CVE-2020-1764}{CVE-2020-1764} &  & \checkmark &  & \\ 
    & 05& Short key length & Low & \href{https://nvd.nist.gov/vuln/detail/CVE-2023-47640}{CVE-2023-47640}$\dagger$ & \checkmark & \checkmark & \checkmark & \checkmark \\ 
     & 06&Static or predictable IV & Med & \href{https://www.cve.org/CVERecord?id=CVE-2024-41260}{CVE-2024-41260}& & \checkmark & \checkmark &  \\
    \midrule
    \multirow{3}{*}{\parbox{2.2cm}{Password-based KDF}} 
    & 07 & Short salt length & Low & \na{} & & \checkmark &  &  \\ 
    & 08 & Predictable salt & Low & \na{} & & \checkmark & &  \\ 
    & 09 & Low hash iterations & Low & \href{https://nvd.nist.gov/vuln/detail/CVE-2023-46233}{CVE-2023-46233}$\dagger$  & & \checkmark & &  \\ 
    \midrule
    \multirow{2}{*}{\parbox{2.2cm}{Transport\\Security}} 
     & 10 & HTTP protocol & High & \href{https://nvd.nist.gov/vuln/detail/CVE-2024-1968}{CVE-2024-1968}$\dagger$ &  & \checkmark &  &  \\
    & 11 & TLS/SSL Issues & High & \href{https://www.cve.org/CVERecord?id=CVE-2024-23656}{CVE-2024-23656} & \checkmark & \checkmark & \checkmark & \checkmark \\ 
    \midrule
    \multirow{2}{*}{\parbox{2.2cm}{Secure Shell}} 
    & 12 &  Insecure SSH suite & High & \href{https://github.com/advisories/GHSA-jj54-5q2m-q7pj}{CVE-2021-32026}& & \checkmark & & \\
   &13 &  No host key validation & High 
    & \href{https://gist.github.com/nyxfqq/33ceaccbc9b05d439a944c2b55fa1c0f}{CVE-2024-41264}& \checkmark  & \checkmark & \checkmark & \\
    \midrule
     \multirow{1}{*}{\parbox{2.2cm}{Token Auth.}}& 14 & No JWT verification & High & \href{https://nvd.nist.gov/vuln/detail/CVE-2024-51744}{CVE-2024-51744}& \checkmark &  &  & \\ 
    \bottomrule \\[-1ex]
\multicolumn{9}{l}{\footnotesize\textit{Note.} $\dagger$ Non-Go advisories; \na{} no associated advisory.}
  \end{tabular}
\end{table*}

\subsubsection{Tool Configuration and Execution}
We select tools that are available and executable, using only their built-in rules. We exclude reports in test files when possible, but do not apply further configuration for filtering alerts. Each tool is executed on every applicable project in our dataset using the mapped rules relevant to that tool.

\subsubsection{Cross-tool Detection Matching}\label{sec:detection-matching}
We map tool-specific rules to our unified taxonomy (\autoref{sec:taxonomy}), then match detections by location and rule. For per-rule agreement, we match by $(\text{project}, \text{file},$ $\text{line}, \text{rule})$, counting each tool once per combination. This captures line-level consensus on specific vulnerability types. For overall agreement, we match by $(\text{project}, \text{file}, \text{line})$ to measure whether tools agree that a line contains \textit{any} vulnerability. Column positions are omitted as they are inconsistently reported across tools.
  
\subsubsection{Evaluation Metrics}\label{sec:metrics}
Without ground truth for tool detections, we rely on \textit{manual analysis} of detections. We also report median wall-clock \textit{execution time} per tool:
\begin{displaymath}
\text{Execution Time}_t = \text{median}_{p \in P}\quad(\text{Time}(t, p))
\end{displaymath}
where $\text{Time}(t, p)$ represents the total execution time for tool $t$ on project $p \in P$ when analyzing all supported rules.
\subsubsection{Experimental Setup}
We conducted our experiments on a server running Ubuntu 22.04, equipped with a 36-core Intel Core i9-10980XE CPU at 3.00 GHz and 125 GB of RAM.
Each tool-project pair runs in a Docker container with 2 CPUs and 9 GB of memory. We use GNU Parallel\footnote{\href{https://www.gnu.org/software/parallel/}{https://www.gnu.org/software/parallel/}} 
 for scheduling.
\section{Qualitative Analysis}
\subsection{Taxonomy of Cryptographic API Misuses}\label{sec:taxonomy}
Our final taxonomy includes \totalnumberofrules distinct misuse issues grouped into six domains. Table~\ref{tab:taxonomy-rules}a) summarizes these issues, their severity, and related advisories. Advisories prefixed \textit{GO} refer to entries in the Go vulnerability database\footnote{\href{https://pkg.go.dev/vuln/}{https://pkg.go.dev/vuln/}}
. Each category reflects real-world patterns observed in production, supported by their implementation in existing tools and associated security advisories.

\subsubsection{Cryptographic Primitives}
This category targets the usage of cryptographic algorithms that are considered fundamentally insecure or deprecated.

\paragraph{Insecure Algorithms} (\#$01$) covers the use of broken cryptographic primitives. For instance, MD5 has been considered insecure since 2004 due to practical collision attacks \cite{hoffman_attacks_2005}. Its continued use in security-critical contexts leads to vulnerabilities. For example, the Go web framework \textit{Beego} used MD5 to generate cache keys (CVE-2024-55885), risking data integrity. Collisions could let attackers overwrite legitimate cache entries, serving malicious responses or enabling unauthorized access.

\paragraph{Insecure Random Number Generation} (\#$02$) covers the use of non-cryptographic randomness in security contexts. Go’s \verb|math/rand| package is predictable and thus insecure \cite{cox_secure_nodate}.
For example, the \textit{Caddy} web server’s \verb|caddy-security| plugin (CVE-2024-21495) used \verb|math/rand| to generate OAuth nonces, allowing attackers to predict values and perform replay attacks with forged requests.

\paragraph{Deprecated Libraries} (\#$03$) relates to the usage of code in the standard library that is considered inefficient, unmaintained, or error-prone. As an example, the \verb|crypto/elliptic|\footnote{\href{https://pkg.go.dev/crypto/elliptic}{https://pkg.go.dev/crypto/elliptic}} package has been deprecated, superseded by \verb|crypto/ecdh|, which provides safer defaults.

\subsubsection{Key Management}
This category collects issues arising from improper generation, management, and usage of cryptographic keys and parameters. Unlike algorithmic weaknesses, where the primitive itself is flawed, these vulnerabilities arise from poor key practices that undermine the otherwise secure cryptographic systems.

\paragraph{Constant/Predictable Keys} (\#$04$) concerns the use of static, hardcoded, or guessable cryptographic keys. In \textit{Kiali} (CVE-2020-1764), a default JWT signing key allowed attackers to forge tokens and gain unauthorized administrative access to the API and logs.

\paragraph{Short Key Length} (\#$05$) addresses cryptographic keys shorter than current security standards. HMAC keys should be at least 128 bits to resist brute-force attacks \cite{sonmez_turan_keyed-hash_2024}.
For example, \textit{DataHub} (CVE-2023-47640) used an 80-bit SHA-1 HMAC for cookie signing, allowing attackers to brute-force the key and gain elevated privileges.

\paragraph{Static or Predictable IVs} (\#$06$) concerns the need for unique, unpredictable initialization vectors (IVs). Static or predictable IVs in block ciphers expose ciphertexts to dictionary attacks \cite{CWE1204GenerationWeak}.
For instance, \textit{NetBird} (CVE-2024-41260) used a hardcoded IV in CBC encryption, causing deterministic outputs that let attackers detect repeated plaintexts or build ciphertext dictionaries.

\subsubsection{Password-based Key Derivation Functions}
This category covers issues in the cryptographic use of passwords, specifically password-based key derivation functions (PBKDFs). Algorithms such as Argon2 \cite{biryukovArgon2NewGeneration2016} and bcrypt \cite{provos_future-adaptable_1999} convert passwords into cryptographic keys through computationally intensive operations, making brute-force guessing significantly harder.

\paragraph{Short Salt Length} (\#$07$) in PBKDFs increases the risk of hash collisions across users, making lookup table attacks more feasible. For a small salt space, attackers can precompute hash values for common passwords and reuse them \cite{provos_future-adaptable_1999}.

\paragraph{Predictable Salts} (\#$08$) addresses vulnerabilities from low-entropy or static salt generation. Hardcoded or predictable salts make hashes guessable via precomputed lookups. For instance, the \textit{Duplicacy} backup tool\footnote{\href{https://github.com/gilbertchen/duplicacy/issues/137}{https://github.com/gilbertchen/duplicacy/issues/137}}
 used a hardcoded salt, undermining salting and enabling lookup-table attacks.

\paragraph{Low Hash Iterations} (\#$09$) in PBKDFs reduces computational cost for attackers, enabling faster brute-force attacks. For instance, a low bcrypt \verb|cost| parameter\footnote{\href{https://pkg.go.dev/golang.org/x/crypto/bcrypt}{https://pkg.go.dev/golang.org/x/crypto/bcrypt}}
 weakens password protection. A notable real-world case is \verb|crypto-js|, which defaulted to a single PBKDF2 iteration (CVE-2023-46233), leaving users exposed to significantly reduced security.

\subsubsection{Transport Security}
This category concerns the secure transmission of data over networks and comprises two rules.

\paragraph{HTTP Protocol} (\#$10$) concerns transmitting sensitive data over unencrypted HTTP, exposing it to eavesdropping and man-in-the-middle attacks. This occurs when applications send credentials or personal data without HTTPS. For example, CVE-2024-1968 affected the Scrapy framework, which failed to remove Authorization headers during HTTPS-to-HTTP redirects, leaking authentication tokens in plaintext.

\paragraph{SSL/TLS Issues} (\#$11$) covers three classes of vulnerabilities in SSL/TLS configuration.
First, outdated protocol versions (SSL 3.0, TLS 1.0, 1.1) contain design flaws \cite{rfc8996}, enabling attacks like POODLE \cite{cisa_poodle}; for example, CVE-2024-23656 in Dex exposed deprecated TLS versions.
Second, insecure cipher suites rely on weak algorithms such as RC4 or lack forward secrecy.
Third, skipping certificate verification allows server impersonation via man-in-the-middle attacks.

\subsubsection{Secure shell (SSH)} issues concerns the configuration and usage of the Go \verb|crypto/ssh|\footnote{\href{https://pkg.go.dev/golang.org/x/crypto/ssh}{https://pkg.go.dev/golang.org/x/crypto/ssh}} package.

\paragraph{Insecure SSH Suite} (\#$12$) covers configurations using outdated ciphers (e.g., CBC mode), which enable plaintext recovery via padding oracle attacks \cite{albrecht2009plaintext}. Although Go’s SSH package disables these by default, users can re-enable them, as seen in \textit{NATS} (CVE-2021-32026), where insecure overrides weakened security.

\paragraph{No Host Key Verification} (\#$13$) arises when SSH key exchange skips host verification, enabling man-in-the-middle attacks. In Go, this typically results from using \verb|ssh.InsecureIgnoreHostKey|. \textit{Casdoor} (CVE-2024-41264) exhibited this flaw, allowing attackers to impersonate trusted servers and intercept sensitive data.
 
\subsubsection{Token-based Authentication}
This category includes one misuse related to JSON Web Tokens (JWTs), widely used for web authentication \cite{jones2015jwt}.

\paragraph{No JWT Verification} (\#$14$) refers to missing signature checks, allowing attackers to forge tokens and access protected resources \cite{jones2015jwt}. In \verb|golang-jwt/jwt|\footnote{\url{https://github.com/golang-jwt/jwt}}
 (CVE-2024-51744), unclear error-handling documentation led developers to accept invalid tokens.

\subsection{Tool Characteristics}
We compare key characteristics distinguishing the tools in our study. \autoref{tab:tool-features} summarizes their availability, configurability, and analytical capabilities.

\paragraph{Availability \& Open-source}
\newcommand{\rotlabel}[1]{\rotatebox[origin=c]{20}{#1}}
\renewcommand{\arraystretch}{1} 
\definecolor{darkred}{rgb}{0.6,0,0}
\begin{table}[bt]
\caption{Tool properties for misuse detection.}
  \label{tab:tool-features}
  \centering
  \scriptsize
\begin{tabular}{lC{0.6cm}C{0.6cm}C{0.6cm}C{0.6cm}C{0.6cm}C{0.6cm}}
  \toprule
  \textbf{Tool} & \rotlabel{Available} & \rotlabel{Open Source} & \rotlabel{Rule Selection} & \rotlabel{Column Prec.} &\rotlabel{Severity} & \rotlabel{Confidence} \\
  \midrule
  \textbf{CodeQL} & \checkmark & \checkmark & \checkmark & \checkmark  & \checkmark &  \textcolor{darkred}{\ding{55}} \\
  \textbf{Gopher} & \checkmark & \textcolor{darkred}{\ding{55}} & \checkmark & \textcolor{darkred}{\ding{55}} &  \textcolor{darkred}{\ding{55}} &  \textcolor{darkred}{\ding{55}} \\
  \textbf{Gosec} & \checkmark & \checkmark & \checkmark & \checkmark  & \checkmark & \checkmark \\
  \textbf{Snyk} & \checkmark & \textcolor{darkred}{\ding{55}} & \checkmark & \checkmark  & \checkmark & \textcolor{darkred}{\ding{55}} \\
  \bottomrule
\end{tabular}
\end{table}

All four tools are publicly available. \CodeQL and \Gosec are fully open-source\footnote{\href{https://opensource.org/osd}{https://opensource.org/osd}}, allowing modification and redistribution. In contrast, \SnykCode and \Gopher are closed-source: \SnykCode as a commercial product and \Gopher as a research binary-only.

\paragraph{Custom Rules}
All tools support adding new detection rules, but with varying flexibility. \CodeQL and \Gosec offer fine-grained control, allowing inclusion/exclusion of specific rules for targeted analysis. \Gopher and \SnykCode lack this suppression feature, forcing users to apply their full rule sets.

\paragraph{Detection Output Characteristics}
The tools differ in how precisely they report findings (granularity, severity, confidence).
Most tools (\CodeQL, \Gosec, \SnykCode) report at column-level precision, while \Gopher reports only by line. All except \Gopher assign severity levels, and only \Gosec includes confidence ratings, helping users prioritize results.

\begin{center}
\begin{findingbox}
\textbf{Finding 1} The tools vary in availability, configurability, and detection precision. 
An ideal tool offers rule selection, column-level accuracy, and severity and confidence ratings. Only \textit{\Gosec} meets all these criteria.
\end{findingbox}
\end{center}

\subsection{Detection Capabilities}
We analyze each tool’s coverage of cryptographic API misuse issues and their detection methods, assessing how comprehensively they address our taxonomy (\autoref{tab:taxonomy-rules}b).

\paragraph{\CodeQL} detects five misuse rules, emphasizing high-severity issues (three of five).
It uniquely identifies unverified JWTs in \texttt{golang-jwt/jwt} (\#$14$). \CodeQL operates in two stages: first, it converts the codebase into a fact database capturing syntactic and semantic relationships \cite{CodeQLQueryHelp2025}; second, QL queries analyze this data, often through data-flow reasoning. 
For instance, Rule \#$02$ (shown in \lstref{lst:codeql_randomness}) models insecure randomness as a taint-analysis problem, tracking flows from \verb|math/rand| to \verb|crypto| APIs while excluding benign cases.

\begin{figure}[tb]
    \centering
\begin{lstlisting}[
    caption={Simplified Rule 02 (\textit{CWE-338}) detection in \CodeQL.},
    label={lst:codeql_randomness},      
    language=QL,
    breaklines=true, 
    frame=lines,                   
    numbers=left,                  
    basicstyle=\scriptsize\ttfamily,
    breakatwhitespace=true,
    columns=fullflexible,  
    ]
class InsecureRandomSource extends Source, DataFlow::CallNode {
  InsecureRandomSource() { getTarget().getPackage().getPath() = "math/rand" }}

class CryptographicSink extends Sink {
  CryptographicSink() {
    exists(Function fn, string pkg, string name |
      pkg.regexpMatch("crypto/.*") and not (pkg="crypto/rand" and name="Read") and
      exists(DataFlow::CallNode c | c.getTarget()=fn and this=c.getAnArgument()))
  } }
\end{lstlisting}
    \Description{Listing of a simplified \CodeQL rule for detecting insecure randomness: it defines sources from math/rand and sinks in crypto APIs, showing how taint flows are tracked to flag misuse.}
\end{figure}

\paragraph{\Gopher} covers 13 of the \totalnumberofrules\ rules (excluding JWT authentication). It targets multiple Go standard library APIs. \Gopher applies taint tracking on the SSA representation \cite{zhangGopherHighPrecisionDeepDive2024a} and supports dynamically updatable rules. For example, when encountering an indirect reference to a cryptographic primitive (e.g., \verb|rsa.GenerateKey|), it can add new rules to detect misuses that do not directly invoke known APIs.

\paragraph{\Gosec} covers six rules across four categories. Like \Gopher, it uses SSA analysis for certain checks (e.g., short keys, static IVs) \cite{gosec}, while others rely on simpler AST inspection. For instance, insecure randomness (\#$02$) is detected by matching function calls against a blacklist.

\paragraph{\SnykCode} supports four rules overlapping with \, including insecure algorithms (\#$01$) and SSL/TLS misconfigurations (\#$11$). It applies static analysis enhanced by an “AI engine” \cite{GoSnykUser2025}, though implementation details remain proprietary.

\begin{center}
\begin{findingbox}
\textbf{Finding 2} Coverage ranges from 4 (\textit{\SnykCode}) to 13 (\textit{\Gopher}) of 14 patterns. Primary detection mode is taint tracking over intermediate representations.
\end{findingbox}
\end{center}

\section{Quantitative Analysis}
We examine how effectively the 4 considered tools that are available (\CodeQL, \Gopher, \Gosec, \SnykCode) detect cryptographic API issues across our dataset of \datasetsize open-source projects.

\subsection{Detection Prevalence}

\begin{table}[b]
  \centering
  \footnotesize
  \caption{Tool execution outcomes across the dataset.}
  \label{tab:execution-results}
\begin{tabular}{l C{0.9cm} C{0.9cm} C{0.9cm} C{0.9cm}}
  \toprule
        & \textbf{CodeQL} & \textbf{Gopher} & \textbf{Gosec} & \textbf{Snyk} \\
  \midrule
  Projects analyzed (\%)            & 91.8 & 77.1 & 100  & 100  \\
  With findings (\% of analyzed)    & 30.9 & 64.8 & 84.5 & 92.7 \\
  With findings (\% of dataset)     & 28.4 & 50.0 & 84.5 & 92.7 \\
  \bottomrule
\end{tabular}
\end{table}

\subsubsection{Reliability and Coverage}
\autoref{tab:execution-results} shows variability in tool reliability and coverage. 
\SnykCode and \Gosec analyze all projects, while \Gopher fails on 75 (22.9\%) 
and \CodeQL on 27 (8.2\%).

Detection coverage varies widely: \SnykCode flags 92.7\% of all projects, \Gosec 84.5\%, \Gopher 50.0\%, and \CodeQL 28.4\%. Most projects are flagged by at least one tool.

\subsubsection{Detection Patterns}
\autoref{tab:tool-rule-matrix} summarizes detection magnitude per tool and misuse category across the dataset. For instance, \CodeQL reports 14 instances of insecure PRNG usage. Missing support for a rule is indicated by a ``--''. 

\begin{table}[tb]
  \caption{Detection counts per rule across analysis tools.}
  \label{tab:tool-rule-matrix}
  \centering
  \footnotesize
  \renewcommand{\arraystretch}{1.2}
  \begin{tabular}{
    p{0.4cm}
    >{\raggedright\arraybackslash}p{2.5cm}
    C{0.7cm} C{0.6cm} C{0.6cm} C{0.6cm} C{0.6cm}
  }
    \toprule
    \multirow{2}{*}{\textbf{ID}} &
    \multirow{2}{*}{\textbf{Description}} &
    \rotlabel{\textbf{CodeQL}} &
    \rotlabel{\textbf{Gopher}} &
    \rotlabel{\textbf{Gosec}} &
    \rotlabel{\textbf{Snyk}} &
    \rotlabel{\textbf{Total}} \\
    \midrule
    \addlinespace[0.5ex]
    01 & Insecure algorithms & \none{} & \cellcolor[rgb]{0.79,0.82,0.89}498 & \cellcolor[rgb]{0.75,0.80,0.87}584 & \cellcolor[rgb]{0.50,0.59,0.71}1259 & 2341 \\
    02 & Crypto insecure PRNG & \cellcolor[rgb]{0.97,0.98,1.00}14 & 0 & \cellcolor[rgb]{0.03,0.19,0.42}\textcolor{white}{2517} & \cellcolor[rgb]{0.97,0.98,1.00}17 & 2548 \\
    03 & Deprecated Go function & \none{} & \cellcolor[rgb]{0.93,0.95,0.98}99 & \none{} & \none{} & 99 \\
    04 & Constant/predictable key & \none{} & \cellcolor[rgb]{0.96,0.97,0.99}44 & \none{} & \none{} & 44 \\
    05 & Short key length & \cellcolor[rgb]{0.97,0.98,1.00}11 & \cellcolor[rgb]{0.97,0.98,1.00}12 & \cellcolor[rgb]{0.97,0.98,1.00}7 & \cellcolor[rgb]{0.97,0.98,1.00}5 & 35 \\
    06 & Static/predictable IV & \none{} & \cellcolor[rgb]{0.97,0.98,1.00}5 & \cellcolor[rgb]{0.97,0.98,1.00}11 & \none{} & 16 \\
    07 & Short salt length & \none{} & \cellcolor[rgb]{0.97,0.98,1.00}18 & \none{} & \none{} & 18 \\
    08 & Predictable salt & \none{} & \cellcolor[rgb]{0.97,0.98,1.00}13 & \none{} & \none{} & 13 \\
    09 & Low hash iterations & \none{} & \cellcolor[rgb]{0.95,0.96,0.99}59 & \none{} & \none{} & 59 \\
    10 & HTTP protocol & \none{} & \cellcolor[rgb]{0.90,0.92,0.96}197 & \none{} & \none{} & 197 \\
    11 & TLS/SSL Issues & \cellcolor[rgb]{0.89,0.92,0.95}210 & \cellcolor[rgb]{0.86,0.89,0.93}300 & \cellcolor[rgb]{0.58,0.65,0.76}1049 & \cellcolor[rgb]{0.79,0.83,0.89}474 & 2033 \\
    12 & Insecure SSH suite & \none{} & \cellcolor[rgb]{0.97,0.98,1.00}5 & \none{} & \none{} & 5 \\
    13 & No host key validation & \cellcolor[rgb]{0.97,0.98,1.00}15 & \cellcolor[rgb]{0.97,0.98,1.00}16 & \cellcolor[rgb]{0.96,0.97,0.99}28 & \none{} & 59 \\
    14 & No JWT verification & \cellcolor[rgb]{0.97,0.98,1.00}6 & \none{} & \none{} & \none{} & 6 \\
    \midrule
    & All & 256 & 1266 & 4196 & 1755 & 7473\\
    \bottomrule
  \end{tabular}
\end{table}
\newcommand*{\rulehigh}[0]{\cellcolor[gray]{0.9}}

Detection varies widely, even within the same category (e.g., \#11). Such discrepancies stem from differing definitions of what constitutes a misuse. For example, in rule \#02 (insecure PRNGs), \Gopher, though claiming support, reports none, while \Gosec flags 2,517, \CodeQL 14, and \SnykCode 17. While it is known that assumptions may differ, having such differences makes it very hard for researchers and practitioners to compare performance. 
\Gopher shows the broadest coverage, uniquely detecting issues for rules \#03, \#04, \#07–\#10, and \#12. However, rules \#03 (deprecation) and \#10 (HTTP) yield excessive warnings, suggesting overly aggressive heuristics.

\SnykCode detects four rule types, particularly flagging insecure algorithms (\#01) 1,259 times, mostly involving MD5 and SHA-1 usage. \Gosec also reports these, but only at import statements, producing fewer findings.

Manual analysis shows that the considered tools do not distinguish whether these usages are security relevant or not. For instance, \Gopher flags all occurrences of \verb|chacha20.NewUnauth| \verb|enticatedCipher|, even when the output is subsequently authenticated (e.g., with Ed25519). 

\begin{figure}
    \centering
    \includegraphics[width=0.60\columnwidth]{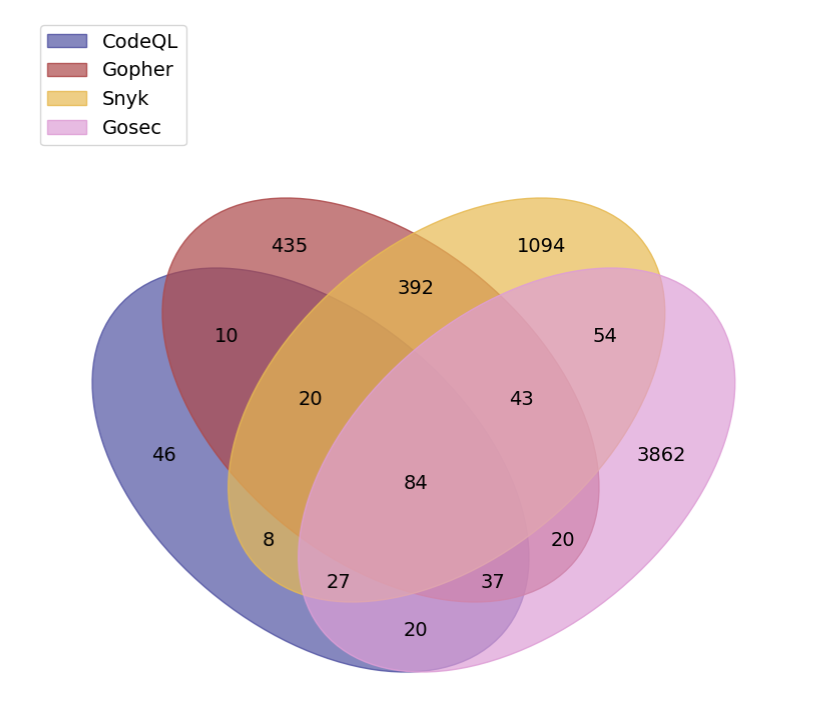}
    \caption{Detection agreement across all rules shows that many detections are tool-unique.}
    \label{fig:venn-overall}
    \Description{Venn diagram comparing tool detections across all rules; each tool has many unique detections with smaller overlaps, indicating limited agreement.}
\end{figure}
\begin{center}
\begin{findingbox}
\textbf{Finding 3} \SnykCode and \Gosec run reliably on all projects; \Gopher fails often yet detects the widest range, sometimes excessively. Large detection discrepancies reveal inconsistent assumptions complicating comparison.
\end{findingbox}
\end{center}
\begin{figure*}[t]
    \centering
    \begin{subfigure}{0.27\linewidth}
        \centering
        \includegraphics[width=1\linewidth]{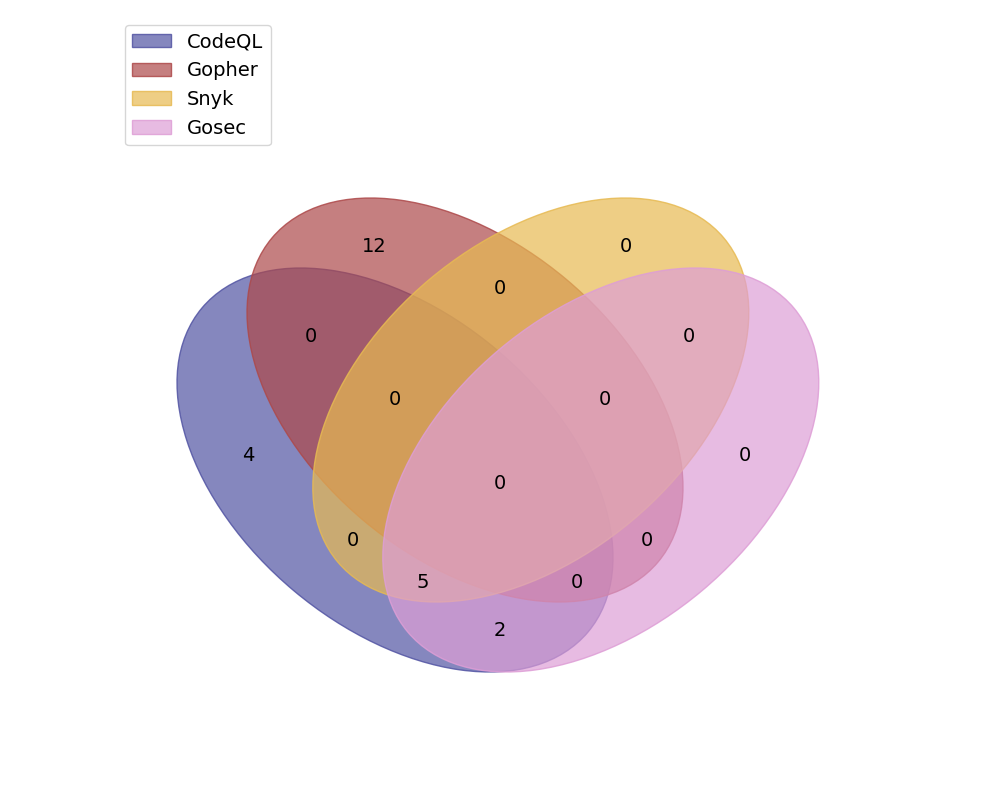}
        \caption{Rule 05 (short key length)}
        \label{fig:venn-cr-05}
    \end{subfigure}
    \hfill
    \begin{subfigure}{0.27\linewidth}
        \centering
        \includegraphics[width=1\linewidth]{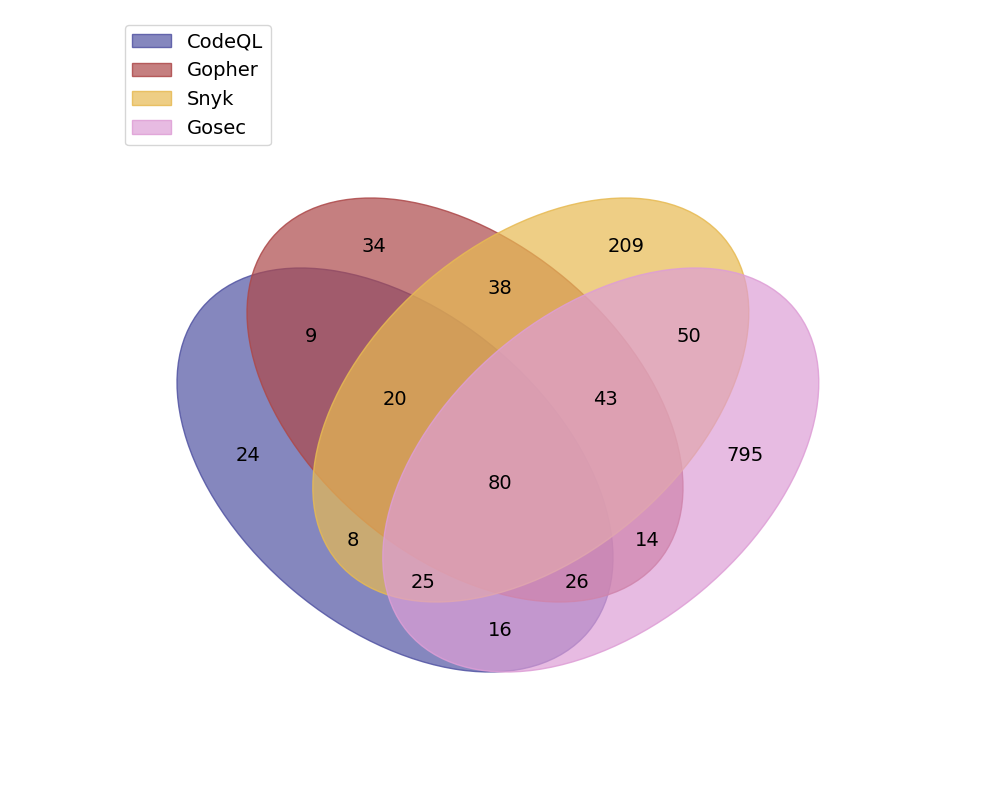}
        \caption{Rule 11 (SSL/TLS issues)}
        \label{fig:venn-cr-11}
    \end{subfigure}
    \hfill
    \begin{subfigure}{0.29\linewidth}
        \centering
        \includegraphics[width=1\linewidth]{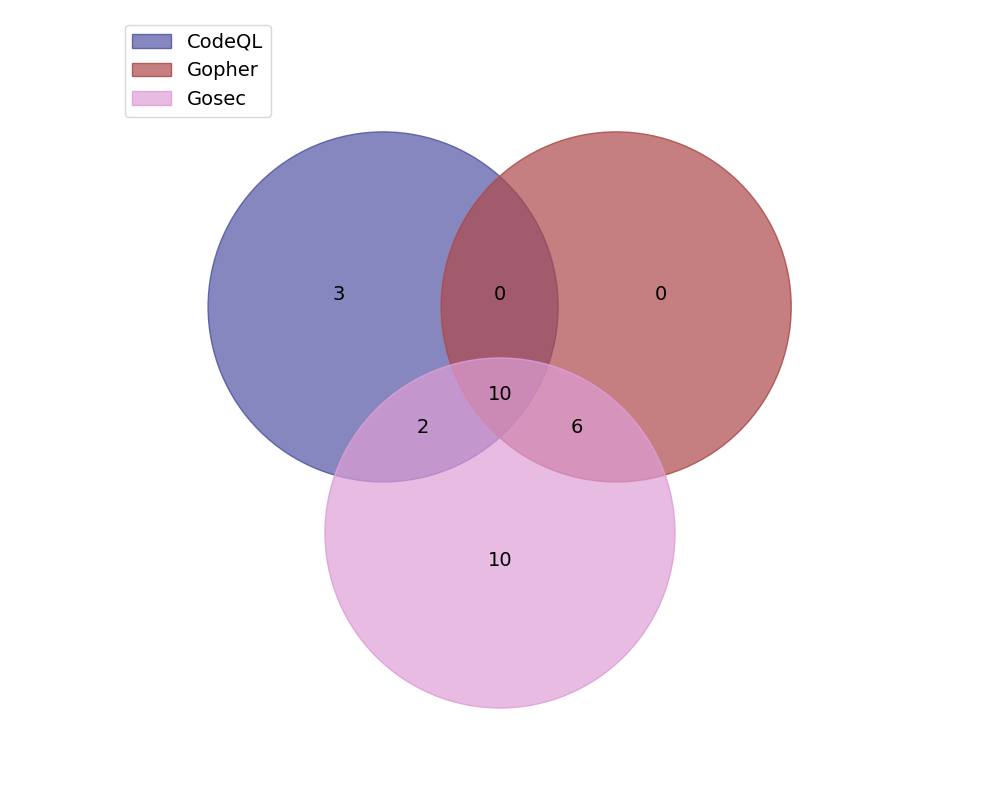}
        \caption{Rule 13 (No key validation)}
        \label{fig:venn-cr-13}
    \end{subfigure}
    \caption{Tool detection overlaps for 3 representative rules. Rule 05 is extreme in the lack of agreement. Rule 11 shows that many detections can be separately confirmed by 2 tools. Rule 13 illustrates rare detections, appropriate for manual analysis.}
    \label{fig:rule-overlaps}
    \Description{Three side-by-side Venn diagrams showing overlap among the four tools for Rule 05, Rule 11, and Rule 13; Rule 05 has minimal overlap, Rule 11 has moderate overlap, and Rule 13 has sparse detections.}
\end{figure*}
\subsection{Tool Consensus}
We measure agreement at the line level, and for rule‑specific analysis, require the same rule at the same line. High overlap suggests stronger confidence, while tool‑specific detections may reflect unique capabilities or false positives; see \autoref{sec:detection-matching}.

\subsubsection{Aggregate Agreement}
Across \totalnumberofrules\ misuse types, we identify 6,152 unique detections. \Gosec accounts for most (67.4\%), followed by \SnykCode (28.0\%), \Gopher (16.9\%), and \CodeQL (4.1\%) (\autoref{fig:venn-overall}). Most findings are tool-specific: \Gosec reports 92.6\% unique locations, \SnykCode 63.5\%, suggesting complementary techniques but also false positive risk. Conversely, 81.7\% of \CodeQL alerts overlap with others. Only 1.3\% of detections are shared by all tools, underscoring fragmented, immature detection in Go.


\subsubsection{Rule-Specific Agreement}
We examine three representative rules with the highest tool coverage (\#05, \#11, \#13) to reveal finer-grained consensus patterns (\autoref{fig:rule-overlaps}).
\paragraph{Short Key Lengths} (\#$05$) is a covered by all four tools. \autoref{fig:venn-cr-05} is the exhaustive analysis of the consensus over all flagged alarms. \Gopher reports 11 unique detections missed by others, while \CodeQL, \Gosec, and \SnykCode agree on 5 locations not flagged by \Gopher.
During manual analysis we observe that 4 of 5 issues flagged by all 3 tools, except \Gopher, are security-irrelevant; they occur in mock and example code and are unlikely to end up in production code. There is one relevant detection that \Gopher misses concerning an RSA key length of 1024, representing a false negative. 


\Gopher’s 11 unique detections reflect a broader interpretation of the rule. Six occur in one project, where it flags unkeyed hashes such as \verb|blake2b.New512(nil)|, which are intended for hashing rather than MACs\footnote{\href{https://pkg.go.dev/golang.org/x/crypto/blake2b}{https://pkg.go.dev/golang.org/x/crypto/blake2b}}. Since static analysis cannot infer context, \Gopher conservatively flags these cases.
In another project, \Gopher identifies low-entropy HMAC usage through a wrapped function call. Although this instance is used for file fingerprinting and not security-relevant, it demonstrates how dynamic rule updates allow \Gopher to detect issues missed by other tools.
Overall, for this rule, \Gopher provides the most advanced detection. With improved robustness (\autoref{tab:execution-results}), it could become a highly effective tool for identifying insecure code patterns.

\paragraph{SSL/TLS Issues} (\#11) yield 1,391 line-level detections, with overlaps shown in \autoref{fig:venn-cr-11}. All tools agree on 80 detections, while \Gosec and \SnykCode dominate unique findings with 795 and 209, respectively. We randomly sampled five detections from each set.  

\Gosec mainly flags TLS version settings and disabled certificate validation via \verb|InsecureSkipVerify|. Three of five samples are false positives where TLS is properly configured (\(\geq\) 1.2), or validation is not actually disabled, suggesting limited sensitivity to configuration values. The remaining two cases are insecure only if developers explicitly enable risky options.  

\SnykCode shows no false positives among its five samples: three correctly identify intentional insecure configurations, and two point to potentially security-relevant issues, indicating stronger contextual awareness than \Gosec. For the five detections shared by all tools, all concern \verb|InsecureSkipVerify| and are true positives: four clearly disable verification, and one reimplements it through a custom method.

\paragraph{SSH Host Key Validation} (\#13) issues are reported by \CodeQL, \Gopher, and \Gosec. These detections focus on how the SSH\\\verb|HostKeyCallback| is implemented. Across tools, there is agreement on 10 alerts, with \CodeQL identifying 3 unique cases and \Gosec 10. We sampled five alerts from each tool’s unique findings.  

\Gosec behaves similarly to rule \#05: it detects true positives of the well-known \verb|InsecureIgnoreHostKey| pattern (1) but also includes a testing file (1) and misses contextual cues such as custom callbacks (1) or explicitly benign configurations (2).  

Samples where all three tools overlap (five alerts) show higher precision, with all being true positives. One of these is intentionally configurable for contexts where no known hosts are available.  



\CodeQL stands out by detecting issues beyond the basic\\\verb|Insecure| \verb|Ignore|\verb|HostKey| pattern. 
It also identifies cases where custom SSH callbacks disable host verification by returning \texttt{nil}, demonstrating its capacity to analyze higher-order functions and complex data flows.

\begin{center}
\begin{findingbox}
\textbf{Finding 4} Detection profiles vary by type: \Gopher offers broad coverage, \SnykCode context-aware precision, \CodeQL unique SSH detection (\#13), while \Gosec produces most false positives. Cross-tool agreement correlates with true positives, suggesting ensemble approaches.
\end{findingbox}
\end{center}
\subsection{Execution Time}
\autoref{tab:median-time} reports median execution time per project, which matters for feasibility in CI and for scaling to large corpora.\CodeQL has the highest overhead due to database creation (144s), for a median total of 219s. \SnykCode is fastest at 16s even when running all available rules. \Gosec follows at 29s, while \Gopher is slower at 63s, reflecting its broader rule coverage.
\begin{center}

\begin{findingbox}
\textbf{Finding 5} \CodeQL is the most time-consuming tool, while \SnykCode runs 13$\times$ faster on median.
\end{findingbox}

\end{center}
\begin{table}[h]
  \centering
  \caption{Median time for setup and analysis per project.}
  \label{tab:median-time}
  \footnotesize
  \renewcommand{\arraystretch}{0.95}
  \begin{tabular}{p{1.6cm}C{1.2cm}C{1.2cm}C{1.2cm}C{1.2cm}}
    \toprule
    & CodeQL & Gopher & Gosec & Snyk* \\
    \midrule
    Setup (s)    & 144 & \na{} & \na{} & \na{} \\
    Analysis (s) & 70  & 63    & 29    & 16    \\
    \bottomrule\\[-0.95ex]
    \multicolumn{5}{l}{\footnotesize\textit{Note.} *For all available Snyk rules, including irrelevant ones.}
  \end{tabular}
\end{table}

\section{Threats to Validity}
\emph{External Validity.} Our 2021 dataset may not capture newer Go projects or emerging misuse patterns, though this is partly mitigated by focusing on long-standing, popular projects.

\emph{Internal Validity.} Mapping tool-specific rules to consolidated misuse categories involved subjective judgment, which may affect conclusions about tool agreement and comparative performance.

\emph{Construct Validity.} Our taxonomy reflects the scope of evaluated tools and may not cover all real-world Go misuses, but it is representative of current automated detection capabilities.
\section{Related Work}
Cryptographic misuse detection has been extensively studied in the Java ecosystem, with several works conducting comparative evaluations of static analysis tools. 

Zhang et al. \cite{zhangAutomaticDetectionJava2023} evaluate six tools (CogniCrypt, CryptoGuard, CryptoTutor, FindSecBugs, SonarQube, Xanitizer) across three Java benchmarks and show that developers often reject tool alerts due to low precision.
Ami et al. \cite{amiWhyCryptodetectorsFail2022} propose MASC, a mutation-based framework for testing detectors under realistic conditions, revealing key practical limitations.
Chen et al. \cite{chenPreciseReportingCryptographic2024} analyze false positives and ineffective true positives in CogniCrypt$\text{SAST}$, CryptoGuard, and CryptoREX, providing recommendations for improving precision.
Wickert et al. \cite{wickertFixNotFix2023} study CogniCrypt$\text{SAST}$ alerts in 210 enterprise projects and find most misuses to be high severity, underscoring their security relevance.

Motivated by recent advances in LLMs, several studies compare them with traditional static analysis for cryptographic misuse detection. Masood and Martin \cite{masoodStaticToolsEvaluating2024} benchmark GPT-4o-mini against CryptoGuard, CogniCrypt, CogniCrypt$_{\text{SAST}}$, and \SnykCode in Java. Xia et al. \cite{xiaExploringAutomaticCryptographic2024} evaluate five LLMs on two Java benchmarks, while Firouzi et al. \cite{firouziChatGPTsPotentialCryptography2024} use prompt engineering with ChatGPT to compare it with CryptoGuard. Overall, LLMs show broader detection coverage but higher false positive rates.

Cryptographic misuse detection has also been explored in Python. Wickert et al. \cite{wickert2021python} present LICMA, while Frantz et al. \cite{frantzMethodsBenchmarkDetecting2024} introduce PyCryptoBench and evaluate Cryptolation against three industry tools on 940 Python projects. Guo et al. \cite{guo2024cryptopyt} propose CryptoPyt and compare it with five tools across two benchmarks. These works highlight that results from Java do not generalize directly, emphasizing the need for language-specific approaches.

Finally, we review prior work on cryptographic misuse detection in Go. Li et al. introduce CryptoGo, the first Go-specific tool, covering 12 misuse types and uncovering issues across all categories in 120 open-source projects. Building on this, Zhang et al. \cite{zhangGopherHighPrecisionDeepDive2024a} propose \Gopher, improving CryptoGo’s detection, and compare both tools to 145 popular Go projects. Their study, however, excludes other Go detectors. We address this gap with the first large-scale comparative evaluation of cryptographic misuse detection tools on real-world, security-relevant Go codebases.

\section{Conclusion}
This paper presented the first systematic study of cryptographic API misuse detection in Go. We evaluated four tools (\CodeQL, \Gopher, \Gosec, and \SnykCode) across 14 rule classes, reporting 7,473 misuses in 328 open-source projects.
Our results reveal major gaps in coverage, precision, and consensus among tools; overlaps are higher-precision, making ensembles a practical near-term strategy while highlighting the need for standardized detection. Future work should combine tool strengths to better support Go security engineers.

\begin{acks}
This work was supported by the Centre for Cyber Defence and Information Security (CDIS), the WASP program funded by Knut and Alice Wallenberg Foundation, and by the Swedish Foundation for Strategic Research (SSF). Some computation was enabled by resources provided by the National Academic Infrastructure for Supercomputing in Sweden (NAISS).
\end{acks}

\bibliographystyle{ACM-Reference-Format}
\bibliography{references}

@misc{CVE-2025-66491,
  author       = {{National Institute of Standards and Technology}},
  title        = {{NVD} - {CVE}-2025-66491},
  howpublished = {\url{https://nvd.nist.gov/vuln/detail/CVE-2025-66491}},
  note         = {Accessed: 2026-01-25},
  year         = {2025},
  publisher    = {National Vulnerability Database (NVD), NIST}
}

@misc{cisa_poodle,
    author = {{CISA}},
    title = {{SSL 3.0 POODLE Attack}},
    year = {2014},
    month = {October},
    day = {17},
    howpublished = {\url{https://www.cisa.gov/news-events/alerts/2014/10/17/ssl-30-protocol-vulnerability-and-poodle-attack}},
    note = {Accessed: 2025-05-30},
    organization = {Cybersecurity and Infrastructure Security Agency}
}

@misc{rfc8996,
    author = {IETF},
    title = {Deprecating TLS 1.0 and TLS 1.1},
    howpublished = {RFC 8996},
    year = {2021},
    month = {March},
    url = {https://datatracker.ietf.org/doc/rfc8996/}
}

@misc{jones2015jwt,
  title = {JSON Web Token (JWT)},
  author = {Jones, Michael and Bradley, John and Sakimura, Nat},
  year = {2015},
  note = {\url{https://datatracker.ietf.org/doc/html/rfc7519}},
  howpublished = {RFC 7519}
}

@inproceedings{albrecht2009plaintext,
  title={Plaintext recovery attacks against SSH},
  author={Albrecht, Martin R and Paterson, Kenneth G and Watson, Gaven J},
  booktitle={2009 30th IEEE Symposium on Security and Privacy},
  pages={16--26},
  year={2009},
  organization={IEEE}
}

@inproceedings{piccolboni2021crylogger,
  title={Crylogger: Detecting crypto misuses dynamically},
  author={Piccolboni, Luca and Di Guglielmo, Giuseppe and Carloni, Luca P and Sethumadhavan, Simha},
  booktitle={2021 IEEE Symposium on Security and Privacy (SP)},
  pages={1972--1989},
  year={2021},
  organization={IEEE}
}

@misc{goBypassCVE,
  author = {{The Go Team}},
  title = {{x/crypto/ssh: misuse of ServerConfig.PublicKeyCallback may cause authorization bypass}},
  howpublished = {\url{https://github.com/golang/go/issues/70779}},
  year = {2024},
  note = {Accessed: 2025-05-30}
}

@misc{CWE1204GenerationWeak,
  author = {MITRE},
  title = {{CWE-1204: Generation of Weak Initialization Vector (IV)}},
  howpublished = {\url{https://cwe.mitre.org/data/definitions/1204.html}},
  note = {Accessed: 2025-05-30}
}

@inproceedings{wickert2021python,
  title={Python crypto misuses in the wild},
  author={Wickert, Anna-Katharina and Baumg{\"a}rtner, Lars and Breitfelder, Florian and Mezini, Mira},
  booktitle={Proceedings of the 15th ACM/IEEE International Symposium on Empirical Software Engineering and Measurement (ESEM)},
  pages={1--6},
  year={2021}
}

@article{afroseEvaluationStaticVulnerability2023,
  title = {Evaluation of {{Static Vulnerability Detection Tools With Java Cryptographic API Benchmarks}}},
  author = {Afrose, Sharmin and Xiao, Ya and Rahaman, Sazzadur and Miller, Barton P. and Yao, Danfeng},
  year = {2023},
  month = feb,
  journal = {IEEE Transactions on Software Engineering},
  volume = {49},
  number = {2},
  pages = {485--497},
  issn = {0098-5589, 1939-3520, 2326-3881},
  doi = {10.1109/TSE.2022.3154717},
  urldate = {2024-12-04},
  copyright = {https://ieeexplore.ieee.org/Xplorehelp/downloads/license-information/IEEE.html},
  langid = {english},
}

@inproceedings{amiWhyCryptodetectorsFail2022,
  title = {Why {{Crypto-detectors Fail}}: {{A Systematic Evaluation}} of {{Cryptographic Misuse Detection Techniques}}},
  shorttitle = {Why {{Crypto-detectors Fail}}},
  booktitle = {2022 {{IEEE Symposium}} on {{Security}} and {{Privacy}} ({{SP}})},
  author = {Ami, Amit Seal and Cooper, Nathan and Kafle, Kaushal and Moran, Kevin and Poshyvanyk, Denys and Nadkarni, Adwait},
  year = {2022},
  month = may,
  pages = {614--631},
  issn = {2375-1207},
  doi = {10.1109/SP46214.2022.9833582},
  urldate = {2024-11-29},
}

@inproceedings{biryukovArgon2NewGeneration2016,
  title = {Argon2: {{New Generation}} of {{Memory-Hard Functions}} for {{Password Hashing}} and {{Other Applications}}},
  shorttitle = {Argon2},
  booktitle = {2016 {{IEEE European Symposium}} on {{Security}} and {{Privacy}} ({{EuroS}}\&{{P}})},
  author = {Biryukov, Alex and Dinu, Daniel and Khovratovich, Dmitry},
  year = {2016},
  month = mar,
  pages = {292--302},
  publisher = {IEEE},
  address = {Saarbrucken},
  doi = {10.1109/EuroSP.2016.31},
  urldate = {2025-05-20},
  isbn = {978-1-5090-1751-5 978-1-5090-1752-2},
  langid = {english},
}

@article{ensafi2015analyzing,
  title={Analyzing the Great Firewall of China over space and time},
  author={Ensafi, Roya and Winter, Philipp and Mueen, Abdullah and Crandall, Jedidiah R},
  journal={Proceedings on privacy enhancing technologies},
  year={2015}
}

@article{mosavi2023detecting,
  title={Detecting Misuse of Security APIs: A Systematic Review},
  author={Mosavi, Seyedehzahra and Islam, Chadni and Babar, Muhammad Ali and Abuadbba, Sharif and Moore, Kristen},
  journal={ACM Computing Surveys},
  year={2023},
  publisher={ACM New York, NY}
}

@unpublished{chenEmpiricalStudyCgo2025,
  title = {{An Empirical Study of Cgo Usage in Go Projects: Distribution, Purposes, Patterns and Critical Issues}},
  author = {Chen, Jinbao and Ding, Boyao and Zhang, Yu and Li, Qingwei and Tang, Fugen},
  year = {2025},
  month = {February},
  eprinttype = {SSRN},
  eprint = {5153961},
  doi = {10.2139/ssrn.5153961},
  url = {https://ssrn.com/abstract=5153961}
}

@inproceedings{chenPreciseReportingCryptographic2024,
  title = {Towards {{Precise Reporting}} of {{Cryptographic Misuses}}},
  booktitle = {Proceedings 2024 {{Network}} and {{Distributed System Security Symposium}}},
  author = {Chen, Yikang and Liu, Yibo and Wu, Ka Lok and Le, Duc V and Chau, Sze Yiu},
  year = {2024},
  publisher = {Internet Society},
  address = {San Diego, CA, USA},
  doi = {10.14722/ndss.2024.241032},
  urldate = {2024-11-29},
  isbn = {978-1-891562-93-8},
  langid = {english},
}

@software{CodeQLQueryHelp2025,
  title = {{CodeQL for Go (v1.1.13)}},
  year = {2025},
  month = may,
  journal = {CodeQL Documentation},
  urldate = {2025-05-01},
  howpublished = {https://codeql.github.com/codeql-query-help/go/},
}

@software{gosec,
  author = {Cosmin Cojocar and Grant Murphy and SecureGo Team},
  title = {{gosec: Go Security Checker (v2.22.4)}},
  year = {2025},
  month = {May},
  url = {https://github.com/securego/gosec},
  publisher = {SecureGo}
}

@inproceedings{firouziChatGPTsPotentialCryptography2024,
  title = {{{ChatGPT}}'s {{Potential}} in {{Cryptography Misuse Detection}}: {{A Comparative Analysis}} with {{Static Analysis Tools}}},
  shorttitle = {{{ChatGPT}}'s {{Potential}} in {{Cryptography Misuse Detection}}},
  booktitle = {Proceedings of the 18th {{ACM}}/{{IEEE International Symposium}} on {{Empirical Software Engineering}} and {{Measurement}}},
  author = {Firouzi, Ehsan and Ghafari, Mohammad and Ebrahimi, Mike},
  year = {2024},
  month = oct,
  pages = {582--588},
  publisher = {ACM},
  address = {Barcelona Spain},
  doi = {10.1145/3674805.3695408},
  urldate = {2025-02-04},
  isbn = {979-8-4007-1047-6},
  langid = {english},
}

@article{frantzMethodsBenchmarkDetecting2024,
  title = {Methods and {{Benchmark}} for {{Detecting Cryptographic API Misuses}} in {{Python}}},
  author = {Frantz, Miles and Xiao, Ya and Pias, Tanmoy Sarkar and Meng, Na and Yao, Danfeng},
  year = {2024},
  month = may,
  journal = {IEEE Transactions on Software Engineering},
  volume = {50},
  number = {5},
  pages = {1118--1129},
  issn = {0098-5589, 1939-3520, 2326-3881},
  doi = {10.1109/TSE.2024.3377182},
  urldate = {2024-11-22},
  copyright = {https://ieeexplore.ieee.org/Xplorehelp/downloads/license-information/IEEE.html},
  langid = {english},
}

@software{GoSnykUser2025,
  title = {{Snyk Code for Go (v1.1297.1)}},
  author = {Snyk},
  year = {2025},
  month = {May},
  url = {https://docs.snyk.io/supported-languages-package-managers-and-frameworks/go},
}

@inproceedings{guo2024cryptopyt,
  title={CryptoPyt: Unraveling Python Cryptographic APIs Misuse with Precise Static Taint Analysis},
  author={Guo, Xiangxin and Jia, Shijie and Lin, Jingqiang and Ma, Yuan and Zheng, Fangyu and Li, Guangzheng and Xu, Bowen and Cheng, Yueqiang and Ji, Kailiang},
  booktitle={2024 Annual Computer Security Applications Conference (ACSAC)},
  pages={1075--1091},
  year={2024},
  organization={IEEE}
}

@inproceedings{krugerCogniCryptSupportingDevelopers2017,
  title={Cognicrypt: Supporting developers in using cryptography},
  author={Kr{\"u}ger, Stefan and Nadi, Sarah and Reif, Michael and Ali, Karim and Mezini, Mira and Bodden, Eric and G{\"o}pfert, Florian and G{\"u}nther, Felix and Weinert, Christian and Demmler, Daniel and others},
  booktitle={2017 32nd IEEE/ACM International Conference on Automated Software Engineering (ASE)},
  pages={931--936},
  year={2017},
  organization={IEEE}
}

@inproceedings{liCryptoGoAutomaticDetection2022,
  title={Cryptogo: Automatic detection of go cryptographic api misuses},
  author={Li, Wenqing and Jia, Shijie and Liu, Limin and Zheng, Fangyu and Ma, Yuan and Lin, Jingqiang},
  booktitle={Proceedings of the 38th Annual Computer Security Applications Conference},
  pages={318--331},
  year={2022}
}

@article{masoodStaticToolsEvaluating2024,
  title={Beyond Static Tools: Evaluating Large Language Models for Cryptographic Misuse Detection},
  author={Masood, Zohaib and Martin, Miguel Vargas},
  journal={arXiv preprint arXiv:2411.09772},
  year={2024}
}

@inproceedings{nadiJumpingHoopsWhy2016,
  title={Jumping through hoops: Why do Java developers struggle with cryptography APIs?},
  author={Nadi, Sarah and Kr{\"u}ger, Stefan and Mezini, Mira and Bodden, Eric},
  booktitle={Proceedings of the 38th International Conference on Software Engineering},
  pages={935--946},
  year={2016}
}

@inproceedings{patnaikUsabilitySmellsAnalysis,
  title={Usability Smells: An Analysis of $\{$Developers’$\}$ Struggle With Crypto Libraries},
  author={Patnaik, Nikhil and Hallett, Joseph and Rashid, Awais},
  booktitle={Fifteenth Symposium on Usable Privacy and Security (SOUPS 2019)},
  pages={245--257},
  year={2019}
}

@inproceedings{rahamanCryptoGuardHighPrecision2019,
  title={Cryptoguard: High precision detection of cryptographic vulnerabilities in massive-sized java projects},
  author={Rahaman, Sazzadur and Xiao, Ya and Afrose, Sharmin and Shaon, Fahad and Tian, Ke and Frantz, Miles and Kantarcioglu, Murat and Yao, Danfeng},
  booktitle={Proceedings of the 2019 ACM SIGSAC Conference on Computer and Communications Security},
  pages={2455--2472},
  year={2019}
}

@inproceedings{wickertFixNotFix2023,
  title={To fix or not to fix: a critical study of crypto-misuses in the wild},
  author={Wickert, Anna-Katharina and Baumg{\"a}rtner, Lars and Schlichtig, Michael and Narasimhan, Krishna and Mezini, Mira},
  booktitle={2022 IEEE International Conference on Trust, Security and Privacy in Computing and Communications (TrustCom)},
  pages={315--322},
  year={2022},
  organization={IEEE}
}

@article{xiaExploringAutomaticCryptographic2024,
  title={Exploring Automatic Cryptographic API Misuse Detection in the Era of LLMs},
  author={Xia, Yifan and Xie, Zichen and Liu, Peiyu and Lu, Kangjie and Liu, Yan and Wang, Wenhai and Ji, Shouling},
  journal={arXiv preprint arXiv:2407.16576},
  year={2024}
}

@article{zhangAutomaticDetectionJava2023,
  title={Automatic detection of Java cryptographic API misuses: Are we there yet?},
  author={Zhang, Ying and Kabir, Md Mahir Asef and Xiao, Ya and Yao, Danfeng and Meng, Na},
  journal={IEEE Transactions on Software Engineering},
  volume={49},
  number={1},
  pages={288--303},
  year={2022},
  publisher={IEEE}
}

@inproceedings{zhangGopherHighPrecisionDeepDive2024a,
  title={Gopher: High-Precision and Deep-Dive Detection of Cryptographic API Misuse in the Go Ecosystem},
  author={Zhang, Yuexi and Li, Bingyu and Lin, Jingqiang and Li, Linghui and Bai, Jiaju and Jia, Shijie and Wu, Qianhong},
  booktitle={Proceedings of the 2024 on ACM SIGSAC Conference on Computer and Communications Security},
  pages={2978--2992},
  year={2024}
}

@misc{hoffman_attacks_2005,
    howpublished = {RFC 4270},
    title = {Attacks on {Cryptographic} {Hashes} in {Internet} {Protocols}},
    url = {https://datatracker.ietf.org/doc/rfc4270},
    author = {Hoffman, Paul E. and Schneier, Bruce},
    month = dec,
    year = {2005},
    doi = {10.17487/RFC4270},
    note = {Num Pages: 12},
}

@misc{cox_secure_nodate,
    title = {Secure {Randomness} in {Go} 1.22},
    howpublished = {https://go.dev/blog/chacha8rand},
    language = {en},
    note = {Accessed: 2025-05-30},
    author = {Cox, Russ and Valsorda, Filippo},
}

@techreport{sonmez_turan_keyed-hash_2024,
    address = {Gaithersburg, MD},
    title = {Keyed-{Hash} {Message} {Authentication} {Code} ({HMAC}): {Specification} of {HMAC} and {Recommendations} for {Message} {Authentication}},
    shorttitle = {Keyed-{Hash} {Message} {Authentication} {Code} ({HMAC})},
    howpublished = {\url{https://nvlpubs.nist.gov/nistpubs/SpecialPublications/NIST.SP.800-224.ipd.pdf}},
    language = {en},
    number = {NIST SP 800-224 ipd},
    urldate = {2025-05-27},
    institution = {National Institute of Standards and Technology},
    author = {Sonmez Turan, Meltem},
    year = {2024},
    doi = {10.6028/NIST.SP.800-224.ipd},
    pages = {NIST SP 800--224 ipd},
}

@inproceedings{provos_future-adaptable_1999,
  title={A future-adaptable password scheme.},
  author={Provos, Niels and Mazieres, David},
  booktitle={USENIX annual technical conference, FREENIX track},
  volume={1999},
  pages={81--91},
  year={1999}
}

\end{document}